\documentclass[reprint,amsmath,amssymb,aps]{revtex4-1}
\usepackage[utf8]{inputenc}
\usepackage{array}
\usepackage{verbatim}
\usepackage{multirow}
\usepackage{amstext}
\usepackage{graphicx}

\makeatletter

\DeclareTextSymbolDefault{\textquotedbl}{T1}
\providecommand{\tabularnewline}{\\}

\usepackage{dcolumn}
\usepackage{bm}

\makeatother

\begin{document}
\title{How to Ensure Logical Consistency of Quantum Theory}
\author{Jindřich Krčmář}
\date{12. 4. 2019}
\begin{abstract}
Predictions of quantum theory have been confirmed experimentally in
the microscopic domain with no known exceptions. This success motivates
physicists to assume universal validity of the theory. To put the
predictions of the quantum theory to the test in the domain of more
complex systems researchers like e.g. Eugene Wigner have proposed
carefully designed Gedankenexperiments revealing unexpected difficulties
of the theory. Daniela Frauchiger and Renato Renner have recently
suggested an extension of the Gedankenexperiment commonly known as
Wigner's friend and arrived at a conclusion that one agent, upon observing
a particular measurement outcome, must conclude that another agent
has predicted the opposite outcome with certainty. Their analysis
shows that quantum theory cannot consistently describe the use of
itself. Here, we will study the mentioned Gedankenexperiments and
introduce an approach leading to consistent predictions of quantum
theory, independent of observer using the theory.
\end{abstract}
\pacs{Valid PACS appear here}

\maketitle


\section{\label{sec:Introduction}Introduction}

Quantum theory has been developed and used over one hundred years
and is nowadays widely accepted in the scientific community \cite{Feynman,Weinberg,domcke2004conical,domcke2011conical,grebenshchikov2007new,gelin2013simulation}.
The enormous success of the quantum theory ranges among many distinct
fields of physics. Yet since the introduction of quantum theory there
have been many discussions about problematic features of the theory
questioning its general validity \cite{WignerMindBody,zurek,FrauchigerRenner}.
In this paper we will study in detail the Wigner's friend \cite{WignerMindBody}
and the Frauchiger-Renner-Wigner \cite{FrauchigerRenner} Gedankenexperiments,
which point out the problems quantum theory faces. We will address
the problems using a unique quantum-mechanical approach.

\begin{figure}
\centering \includegraphics[width=0.48\textwidth]{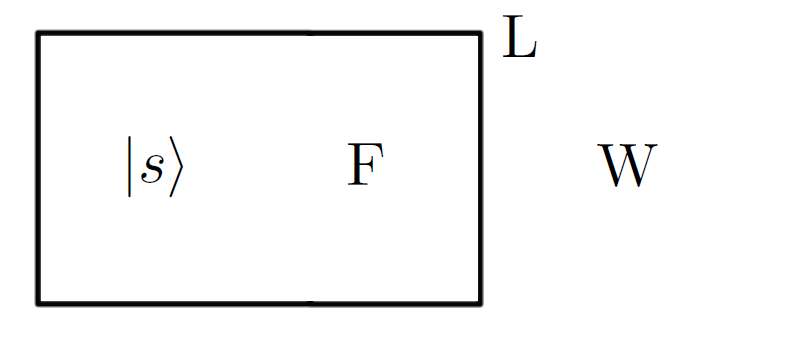} \caption{\label{WignersFriend}Scheme of the Wigner's friend Gedankenexperiment.
Wigner is denoted as W standing outside the laboratory L. The observer
F performs a measurement on a system prepared in a state $|s\rangle=\frac{1}{\sqrt{2}}\left(|\uparrow\rangle+|\downarrow\rangle\right)$.}
\end{figure}

Our first chosen Gedankenexperiment has been proposed by the theoretical
physicist Eugene Wigner in 1961 (famous for his contributions regarding
fundamental symmetry principles in atomic physics \cite{WignerSymmetry}).
In his original work \cite{WignerMindBody} Wigner discusses how the
concept of consciousness enters the quantum theory: “The impression
which one gains at an interaction, called also the result of an observation,
modifies the wave function of the system. The modified wave function
is, furthermore, in general unpredictable before the impression gained
at the interaction has entered our consciousness: it is the entering
of an impression into our consciousness which alters the wave function
because it modifies our appraisal of the probabilities for different
impressions which we expect to receive in the future. It is at this
point that the consciousness enters the theory unavoidably and unalterably.
If one speaks in terms of the wave function, its changes are coupled
with the entering of impressions into our consciousness. If one formulates
the laws of quantum mechanics in terms of probabilities of impressions,
these are ipso facto the primary concepts with which one deals.”.
Wigner as well analyses how sensations of a consciousness depend on
a physical state of the observer: “Let us first specify the question
which is outside the province of physics and chemistry but is an obviously
meaningful (because operationally defined) question: Given the most
complete description of my body (admitting that the concepts used
in this description change as physics develops), what are my sensations?”.
He postulates a thesis about necessary conditions for the consciousness
to arise: “It is very likely that, if certain physico-chemical conditions
are satisfied, a consciousness, that is, the property of having sensations,
arises. This statement will be referred to as our first thesis. The
sensations will be simple and undifferentiated if the physico-chemical
substrate is simple; it will have the miraculous variety and colour
which the poets try to describe if the substrate is as complex and
well organized as a human body. The physico-chemical conditions and
properties of the substrate not only create the consciousness, they
also influence its sensations most profoundly. Does, conversely, the
consciousness influence the physicochemical conditions? In other words,
does the human body deviate from the laws of physics, as gleaned from
the study of inanimate nature?” We will come back to Wigner's first
thesis later on in our discussion. Wigner bases his Gedankenexperiment
on the fact that in the quantum theory the wave function depends on
observer's information. The experiment consists of two observers -
Wigner (W) and his friend (F) (see scheme of the experiment in FIG.
\ref{WignersFriend}). The observer F performs a measurement in an
isolated lab L, let us say a measurement of a spin of a particle,
which is in a normalized state
\begin{equation}
|s\rangle=\frac{1}{\sqrt{2}}\left(|\uparrow\rangle+|\downarrow\rangle\right),\label{1}
\end{equation}
where the states $|\uparrow\rangle$ and $|\downarrow\rangle$ denote
the spin state ``up'' and ``down'' respectively. The result of
his measurement is either an electron spin state $|\uparrow\rangle$
or $|\downarrow\rangle$. From his perspective, the information has
arrived to his consciousness and the state vector has collapsed to
a definite value. From the perspective of the observer W, the situation
looks different. The information about the spin of the electron has
not arrived to W's consciousness and therefore the state is after
the F's measurement still a sum of vectors 
\begin{equation}
\frac{1}{\sqrt{2}}\left(|\uparrow\rangle|F_{\uparrow}\rangle+|\downarrow\rangle|F_{\downarrow}\rangle\right),\label{2}
\end{equation}
where $|F_{\uparrow}\rangle$ and $|F_{\downarrow}\rangle$ describe
the state of the Wigner's friend after measuring the spin state $|\uparrow\rangle$
or $|\downarrow\rangle$ respectively. This experiment demonstrates
that the description of the system is different for both of the observers.
Though the situation may appear “crazy”, this result does not rule
out quantum mechanics as a universally valid theory. Let us rather
conclude at this point that Wigner's friend experiment emphasizes
the importance of the distinct levels of knowledge of the experiment
participating agents. We will analyse the experiment introducing a
different approach to quantum theory and find out that in reality
such experiments should be perceived as observer independent.

\begin{figure}
\centering \includegraphics[scale=0.5]{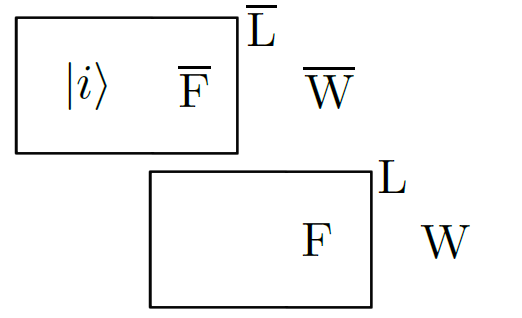} \caption{\label{FRWexperiment} Scheme of the FRW Gedankenexperiment. Agents
$\overline{\text{W}}$ and W stand outside the isolated laboratories
$\overline{\text{L}}$ and L waiting for the measurement results of
their friends $\overline{\text{F}}$ and F. Observer $\overline{\text{F}}$
measures state of a two-level system in state $|i\rangle$ and prepares
another state based on his result for the agent F. After both friends
$\overline{\text{F}}$ and F finish their measurements, the agent
$\overline{\text{W}}$ measures the state in the laboratory $\overline{\text{L}}$
and then the agent W similarly measures the state in the laboratory
L.}
\end{figure}

An extension of the Wigner's friend experiment has been discussed
in a recent publication by Daniela Frauchiger and Renato Renner (FRW
experiment) \cite{FrauchigerRenner}. As you can see in FIG. \ref{FRWexperiment}
we consider two separate laboratories L, $\overline{\text{L}}$ with
two observers F, $\overline{\text{F}}$ performing measurements and
two other observers W, $\overline{\text{W}}$ measuring the state
from the outside of the isolated laboratories respectively.

The experiment is repeated in a loop and proceeds as follows:
\begin{enumerate}
\item Agent $\overline{\text{F}}$ performs a measurement on a two-level
system prepared in initial state 
\begin{equation}
|i\rangle=\frac{1}{\sqrt{3}}|h\rangle+\sqrt{\frac{2}{3}}|t\rangle,\label{3}
\end{equation}
where h stands for ``heads'' and t for ``tails''. Based on the
resulting state vector either $|h\rangle$ or $|t\rangle$, agent
$\overline{\text{F}}$ sets the particle spin to state $|\downarrow\rangle$
or $|\rightarrow\rangle=\frac{1}{\sqrt{2}}\left(|\uparrow\rangle+|\downarrow\rangle\right)$
respectively. The particle is then transferred to the observer F.
\item The agent F measures the spin with respect to the basis $|\uparrow\rangle$
and $|\downarrow\rangle$.
\item In the third step the agent $\overline{\text{W}}$ measures the state
in the laboratory $\overline{\text{L}}$ obtaining its projection
on basis vectors $|\overline{\text{ok}}\rangle=\frac{1}{\sqrt{2}}\left(|h\rangle-|t\rangle\right)$
and $|\overline{\text{fail}}\rangle=\frac{1}{\sqrt{2}}\left(|h\rangle+|t\rangle\right)$.
\item Finally the observer W concludes the round of this experiment by measuring
the state in the laboratory L with respect to the basis $|\text{ok}\rangle=\frac{1}{\sqrt{2}}\left(|\downarrow\rangle-|\uparrow\rangle\right)$
and $|\text{fail}\rangle=\frac{1}{\sqrt{2}}\left(|\downarrow\rangle+|\uparrow\rangle\right)$.
The experimental procedure is repeated until the observers $\overline{\text{W}}$
and W both obtain the results $|\overline{\text{ok}}\rangle$ and
$|\text{ok}\rangle$ respectively.
\end{enumerate}
In the case of Wigner's experiment we used $|\uparrow\rangle|F_{\uparrow}\rangle$
to denote the combined state of the observer F and particle spin. We
have simplified the notation for the more complex FRW experiment for
brevity. Further composition details of the vectors showing the connection
to the agents $\overline{\text{F}}$, F, $\overline{\text{W}}$
and W in the same way would make our notation look unnecessarily more
complicated and interested readers can find it in the original paper
if needed \cite{FrauchigerRenner}. Similarly as in the case of Wigner's
friend experiment, after every measurement the state vector collapses
to a certain basis vector from the perspective of the observer performing
the measurement, but is still perceived as a superposition of states
from the perspective of the agents not participating the measurement.
After a systematic analysis of the experiment, Frauchiger and Renner
have concluded that the observers $\overline{\text{W}}$ and W will
unavoidably come to contradictory predictions using the quantum theory.
According to quantum mechanics the experiment will halt with agents
$\overline{\text{W}}$ and W measuring $|\overline{\text{ok}}\rangle$
and $|\text{ok}\rangle$ respectively at one point despite the fact
that after $\overline{\text{W}}$'s measurement of the state $|\overline{\text{ok}}\rangle$
the agent W's measurement can not result in the state $|\text{ok}\rangle$
according to the same quantum theory. The theory has consequently
been proven to contain contradictory predictions in itself.

\section{\label{sec:Materials}Observer independent approach to quantum mechanics}

Let us come back to Wigner's first thesis: \textquotedbl It is very
likely that, if certain physico-chemical conditions are satisfied,
a consciousness, that is, the property of having sensations, arises.\textquotedbl .
For our discussion we will not attempt to find out the laws of what
exact sensations an observer in a system experiences as a function
of time. For this article, the observation that a substance can have
sensations is sufficient. From reasons going beyond the scope of this
paper, let us assume that any physical system can be divided into
parts, each being treated as an observer from a quantum theory point
of view. The actual sensations of the observers at an arbitrary point
of time, if there are any, can be considered irrelevant for our purposes.
We will assume two theses as a basis of our considerations:
\begin{itemize}
\item \textbf{Thesis 1}: Every system at any time consists of observers,
completely described by their state vectors - independent wave functions.
\item \textbf{Thesis 2}: Collapse of a state vector appears only upon interaction
between two observers. Interaction within parts (particles) of one
observer does not result in a state vector collapse.
\end{itemize}
Independence of the wave functions means that they are defined by
states of distinct particles in the system. We can assume that a particle
belonging to two distinct observers at a time does not exist. In order
to guarantee clarity of our two universally valid theses and their
consequences, let us analyse the two Gedankenexperiments mentioned
before.

\section{\label{sec:Procedure}Wigner's friend Gedankenexperiment}

Due to the Thesis 1, the fundamental description of the experiment
differs from those used in \cite{FrauchigerRenner}. Consequently
analysis of a quantum-mechanical experiment does not start with one
wave function of the whole system, but with independent wave functions
for each observer. We have to realize, which part of the experiment
belongs to which observer. Our relevant observers are Wigner W, his
friend F, the measurement device M and the particle P. We may perhaps
hesitate to accept that the measurement device and even the particle
can be effectively regarded as observers. To become more comfortable
about the assignment, let us point out several crucial facts. Treating
a measurement device as an observer is relatively common in present
publications \cite{ExperimentalRejectionProietti} and does not necessarily
mean that the observer experience complex sensations comparable to
those of a human mind. The sensations depend on a physical state of
the observers and hence we can calmly assume that sensations of a
measurement device are practically non-existent. What matters for
us is the fact that the sensations can potentially arise according
to some presently unknown laws of the process, because from this reason
we treat any substance as an (potentially conscious) observer. For
the particle we should emphasize that we don't claim it is \textquotedbl a
one-particle observer\textquotedbl , but that only one particle belonging
to some observer is relevant for the experiment. Similarly when we
talk about the measurement device - observer we actually mean only
those parts involved in the measurement process. There is an important
possibility that the particle P belongs to the observer M though.
Not \textquotedbl belongs\textquotedbl{} in the sense of being a
physical part of the device, but in the sense of abstract assignment
of the relevant particles to one observer (independent wave function).

Our Thesis 2 determines that the state vector collapse will appear
no matter whether observer W is a part of the measurement or not,
because P and M are two distinct observers. In case of all the parts
of the experiment belong to different agents, the experiment does
not contain any superposition of macroscopic objects irrespective
of whether we describe the experiment from the perspective of Wigner
or any other agent. Which agent's perspective we consider is not important
for the predictions of the quantum theory anymore. We can describe
the initial state in the same way as previously by $|s\rangle$ (see
Eq. 1). The state of the particle after the interaction between the
P and the M would be then either $|\uparrow\rangle\langle\uparrow|s\rangle$
or $|\downarrow\rangle\langle\downarrow|s\rangle$. Here we have not
included the states of the W, F and M, because our description does
not depend on the observer, who uses quantum theory.

Let us show how it is possible to get the superposition as the one
from Wigner's perspective mentioned in the introduction. The reason
is that we don't have a universal way to find out, which parts of
the experiments are distinct observers and which are in reality belonging
to only one observer. The state of the experiment could be prepared
in such a way that the Wigner's friend F, the measurement device M
and the particle P all belong to one observer. The wave-function collapse
will not occur according to the Thesis 2 and the laboratory L keeps
the superposition state $|s\rangle$ until the agent W performs his
measurement. His measurement will collapse the wave function, because
W and observer device M, particle P and friend F are two distinct
observers.

The experiment could be also prepared in such a way that it is a matter
of chance whether the particle belongs to the same observer as the
rest of the lab or not. In this case we would integrate a probability
factor in the description. In our approach it is irrelevant whether
the device M and agent F are macroscopic or microscopic. Macroscopic
objects are not likely to contain only one observer so the Wigner's
scenario would be significantly more complicated to prepare, because
it would be necessary to overcome this difficulty and prepare such
a one-observer macroscopic object. Nevertheless it is in principle
possible and it would cause no difficulty to the theory.

We have considered two possible distinct experimental set ups of the
Wigner's friend scenario leading us to another understanding of the
subject. Other combinations of ``merged'' observers are not relevant
at the moment. To summarize, the relevant description of the experiment
before W's measurement is $|\uparrow\rangle\langle\uparrow|s\rangle$
or $|\downarrow\rangle\langle\downarrow|s\rangle$ in case of the
parts of the experiments are distinct observers. In case of M, P and
F being one observer the system keeps the superposition state $|s\rangle$
until W performs his measurement. We are now ready to discuss the
more complicated case of the FRW experiment.

\section{\label{sec:DataAnalysis}Frauchiger-Renner-Wigner Gedankenexperiment}

For the FRW scenario we simply have more possibilities of observer
assignments to consider. Everything works in an analogous way as in
the previous case. No logical inconsistency is possible, because the
derivation used in the original paper \cite{FrauchigerRenner} does
not apply in our approach. We have a quantum theory that does not
directly depend on the point of view of an observer.

As in the Wigner's friend scenario analyzed above, let us first assume
all the parts of the experiment belong to distinct observers. Then
apart from the initial state $|i\rangle$ (Eq. (\ref{3})), there
is no newly arising superposition of states due to our Thesis 2. The
process of the experiment is defined by action sequence of the four
projection operators $A_{\centerdot}^{\centerdot}$ in TABLE \ref{unicornio-1}
on the initial state. The projection operators describe the interactions
between distinct observers and are therefore denoted as interaction
operators. The probabilities of all possible measurement outcomes
of all the four observers can be expressed as
\begin{equation}
P_{\varnothing}(x,y,z,w)=\left|\langle z|\langle w|A_{w}^{IV}A_{z}^{III}A_{y}^{II}A_{x}^{I}|i\rangle\right|^{2},
\end{equation}
where $|i\rangle$ is defined in Eq. (\ref{3}), the operators $A_{\centerdot}^{\centerdot}$
in TABLE \ref{unicornio-1} and $x,y,z,w$ describe one of the two
possible measurement outcomes of each observer (for example $z$ can
be either $\overline{\text{ok}}$ or $\overline{\text{fail}}$). This
formula allows us to calculate the probabilities of the outcomes using
the definitions of the relevant vectors and operators.

It is possible to prepare the experiment in analogous way as in the
Wigner's friend experiment considering everything involved in the
$\overline{\text{F}}$'s measurement as one observer, which leads
us to an isometry 
\begin{equation}
U^{I}=|h\rangle\rightarrow|h\rangle|\downarrow\rangle,\:|t\rangle\rightarrow|t\rangle|\rightarrow\rangle,
\end{equation}
which will conserve the superposition of states. The probability distribution
for this experiment is then:
\begin{equation}
P_{1}(y,z,w)=\left|\langle z|\langle w|A_{w}^{IV}A_{z}^{III}A_{y}^{II}U^{I}|i\rangle\right|^{2}.
\end{equation}
We can assume the same for the observer F using isometry $U^{II}=|\psi\rangle\rightarrow|\psi\rangle$
(for every state $|\psi\rangle$). Supposing only F's procedure is
done by a single observer we obtain
\begin{equation}
P_{2}(x,z,w)=\left|\langle z|\langle w|A_{w}^{IV}A_{z}^{III}U^{II}A_{x}^{I}|i\rangle\right|^{2}.
\end{equation}
There is also an option that both observers $\overline{\text{F}}$
and F are the only observers involved in their measurement, implying
\begin{equation}
P_{1,2}(z,w)=\left|\langle z|\langle w|A_{w}^{IV}A_{z}^{III}U^{II}U^{I}|i\rangle\right|^{2}.\label{8}
\end{equation}
The subscript $o$ in $P_{o}$ indicates, which measurements do not
cause the wave function collapse due to the Thesis 2 and the one-observer
nature of the system (measurements $1$ and $2$ are done by the observers
$\overline{\text{F}}$ and F respectively). Let us suppose for simplicity
that in the measurements of $\overline{\text{W}}$ and W there are
always many observers involved, so we don't have to consider more
superposition conserving transformations. In the more general situation,
when the probability of a such a setup $o$ is $p_{o}$, we arrive
at the final probability distribution
\begin{equation}
P(x,y,z,w)=p_{\varnothing}P_{\varnothing}(.)+\frac{p_{1}}{2}P_{1}(.)+\frac{p_{2}}{2}P_{2}(.)+\frac{p_{1,2}}{4}P_{1,2}(.),
\end{equation}
where the $(.)$ in $P_{o}(.)$ is an abbreviation for the variables
of the probability distribution. We have included the normalization
factors in the fractions $\frac{p_{o}}{n}$ ($n$ is integer), because
for example the probability distribution $P_{1}$ does not depend
on the variable $x$ (because the particular state $|x\rangle$ does
not occur in this scenario) and would be then calculated twice (analogously
for the other normalization factors).

\begin{table}[h!]
\caption{FRW experimental procedure chronologically ordered starting from the
top row. The column interaction operators defines the projection operators
$A_{\centerdot}^{\centerdot}$ for all the four inter-observer interactions
in the relevant measurement basis.}
\begin{tabular}{|c|c|c|}
\hline 
Observer & Measurement basis & Interaction operators\tabularnewline
\hline 
\multirow{2}{*}{$\overline{\text{F}}$} & \multirow{2}{*}{$|h\rangle$, $|t\rangle$} & $A_{h}^{I}=|h\rangle|\downarrow\rangle\langle h|$,\tabularnewline
 &  & $A_{t}^{I}=|t\rangle|\rightarrow\rangle\langle t|$\tabularnewline
\hline 
\multirow{2}{*}{F} & \multirow{2}{*}{$|\downarrow\rangle$, $|\uparrow\rangle$} & $A_{\downarrow}^{II}=|\downarrow\rangle\langle\downarrow|$,\tabularnewline
 &  & $A_{\uparrow}^{II}=|\uparrow\rangle\langle\uparrow|$\tabularnewline
\hline 
\multirow{2}{*}{$\overline{\text{W}}$} & $|\overline{\text{ok}}\rangle=\frac{1}{\sqrt{2}}\left(|h\rangle-|t\rangle\right)$, & $A_{\overline{\text{ok}}}^{III}=|\overline{\text{ok}}\rangle\langle\overline{\text{ok}}|$,\tabularnewline
 & $|\overline{\text{fail}}\rangle=\frac{1}{\sqrt{2}}\left(|h\rangle+|t\rangle\right)$ & $A_{\overline{\text{fail}}}^{III}=|\overline{\text{fail}}\rangle\langle\overline{\text{fail}}|$\tabularnewline
\hline 
\multirow{2}{*}{W} & $|\text{ok}\rangle=\frac{1}{\sqrt{2}}\left(|\downarrow\rangle-|\uparrow\rangle\right)$, & $A_{\text{ok}}^{IV}=|\text{ok}\rangle\langle\text{ok}|$,\tabularnewline
 & $|\text{fail}\rangle=\frac{1}{\sqrt{2}}\left(|\downarrow\rangle+|\uparrow\rangle\right)$ & $A_{\text{fail}}^{IV}=|\text{fail}\rangle\langle\text{fail}|$\tabularnewline
\hline 
\end{tabular}\label{unicornio-1}
\end{table}
The quantum mechanical description is the same from the point of view
of all observers. Every observer experiences different sensations
during the experiment, but at any time during the experiment there
exist a set of \textquotedbl the facts of the world\textquotedbl{}
\cite{ExperimentalRejectionProietti} in terms of states of the considered
independent wave functions. We obtain the probability $\frac{1}{12}$
of the experiment to halt obtained in \cite{FrauchigerRenner} using
Eq. (\ref{8}) 
\begin{equation}
P_{1,2}(\overline{\text{ok}},\text{ok})=\left|\langle\overline{\text{ok}}|\langle\text{ok}|A_{\text{ok}}^{IV}A_{\overline{\text{ok}}}^{III}U^{II}U^{I}|i\rangle\right|^{2}=\frac{1}{12}.
\end{equation}
This probability is different for the other scenarios described by
the distributions $P_{\varnothing}(x,y,z,w)$, $P_{1}(y,z,w)$ and
$P_{2}(x,z,w)$. This brings us to an interesting conclusion that
experiments of this kind can help us to measure the probabilities
$p_{o}$ and hence studying the observer composition of the experiment.
Similar experiment has been recently performed \cite{ExperimentalRejectionProietti}.

\section{\label{sec:Results} Summary}

We have introduced an approach to quantum mechanics unique in the
way of defining and treating observers involved in the studied system.
The basis of our considerations are the Theses 1 and 2. The Thesis
1 tells us to describe a system as set of independent wave functions
belonging to distinct observers. The Thesis 2 then postulates how
these observers interact. Our analysis of the Wigner's friend and
the FRW Gedankenexperiments has shown us that we have no difficulties
regarding logical consistency of the theory, unlike the approaches
discussed in \cite{FrauchigerRenner}. The experiments of the FRW
type can help us studying the observer composition of the experiment
due to the probability factors $p_{o}$. Similar experiment has been
recently performed \cite{ExperimentalRejectionProietti} and could
be used for further analysis.

\section{\label{sec:Conclusions} Acknowledgments}

My special thanks belong to Lenka Šlapáková, Sergy Grebenshchikov
and Maxim Gelin.


\begin{thebibliography}{10}
\bibitem{Feynman}Feynman, Leyton, Sands, \emph{Lectures on physics
Vol. 3 }(AW, 1964)

\bibitem{Weinberg}Weinberg, S., \emph{The Quantum Theory of Fields}
(Cambridge University Press, 1995)

\bibitem{domcke2004conical} Domcke W, Yarkony D~R and K{ö}ppel
H {\em Conical Intersections: Electronic Structure, Dynamics and
Spectroscopy\/} (World Scientific, 2004)

\bibitem{domcke2011conical} Domcke W, Yarkony D~R and K{ö}ppel
H {\em Conical Intersections: Theory, Computation and Experiment\/}
(World Scientific, 2011)

\bibitem{grebenshchikov2007new} Grebenshchikov S~Y, Qu Z~W, Zhu
H and Schinke R 2007 {\em Physical Chemistry Chemical Physics\/}
\textbf{9} 2044

\bibitem{gelin2013simulation} Gelin M, Tanimura Y and Domcke W 2013
{\em The Journal of Chemical Physics\/} \textbf{139} 214302

\bibitem{WignerMindBody}Wigner, E. P., \emph{Remarks on the mind--body
question. in Symmetries and Reflections, pp. 171--184} (Indiana University
Press, 1967)

\bibitem{zurek}W. H. Zurek 2007 \emph{ArXiv 0707.2832}

\bibitem{FrauchigerRenner}Frauchiger D, Renner R 2018 \emph{Nature
Communications}\textbf{ 9} 3711

\bibitem{WignerSymmetry}Wigner, E. P., \emph{Group Theory and its
Application to the Quantum Mechanics of Atomic Spectra }(Academic
Press, 1959)

\bibitem{ExperimentalRejectionProietti}Proietti M, Pickston A, Graffitti
F, Barrow P, Kundys D, Branciard C, Ringbauer M, Fedrizzi A 2019 \emph{arXiv:1902.05080}
\end{thebibliography}
\end{document}